# Ligand Additivity and Divergent Trends in Two Types of Delocalization Errors from Approximate Density Functional Theory

Yael Cytter[1], Aditya Nandy[1,2], Akash Bajaj[1,3], and Heather J. Kulik[1,*]

[1]Department of Chemical Engineering, Massachusetts Institute of Technology, Cambridge, MA 02139, USA

[2]Department of Chemistry, Massachusetts Institute of Technology, Cambridge, MA 02139, USA

[3]Department of Materials Science and Engineering, Massachusetts Institute of Technology, Cambridge, MA 02139, USA

AUTHOR INFORMATION

**Corresponding Author**

*email: hjkulik@mit.edu, phone: 617-253-4584

ABSTRACT

Despite its widespread use, the predictive accuracy of density functional theory (DFT) is hampered by delocalization errors, especially for correlated systems such as transition-metal complexes. Two complementary tuning strategies have been developed to reduce delocalization error: eliminating the global curvature with respect to charge addition or removal, and computing a linear response Hubbard U as a measure of local curvature at the metal center at fixed charge and applying it to the transition-metal complex in a DFT+U framework. We investigate the relationship between the two measures of delocalization error as we manipulate the ligand field strength by varying the number of strong-field ligands in a series of heteroleptic complexes or by geometrically constraining the metal–ligand bond length in homoleptic octahedral complexes. We show that across these sets of complexes with varying ligand fields, an inverse relationship generally exists between global and local curvatures. We find that effects of ligand substitution on both measures of delocalization are typically additive, but the two quantities seldom coincide. The observation of ligand additivity suggests opportunities for evaluating errors on homoleptic complexes to infer corrections for lower-symmetry complexes.

**TOC GRAPHICS**

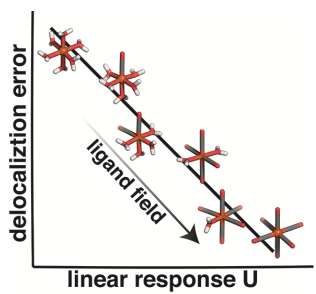

Density functional theory (DFT) is widely used in theoretical studies of transition-metal complexes (TMCs).[1, 2] Common approximations to the exchange-correlation (xc) functional, however, introduce self-interaction errors (SIEs)[2-6], resulting in fundamental errors in the description of dissociation energies[4, 7-10], barrier heights[11], band gaps[12, 13] and electron affinities[14-16] as well as thermodynamic properties.[17] Because the exact energy functional should be piecewise linear with respect to fractional removal or addition of charge[18], the global curvature, i.e., the second derivative of the energy $E$ with respect to charge $q$, is considered a good measure for many-electron self-interaction error also known as delocalization error (DE). Using Koopman's[19] theorem, it was shown the average global curvature can be approximated by the following expression:

$$\left\langle \frac{\partial^2 E}{\partial q^2} \right\rangle = \epsilon_{N+1}^{HOMO} - \epsilon_{N}^{LUMO}, \quad (1)$$

where $\epsilon_{N+1}^{HOMO}$ is the energy of the highest occupied molecular orbital (HOMO) of a system with $N+1$ electrons, and $\epsilon_{N}^{LUMO}$, is the energy of the lowest unoccupied molecular orbital (LUMO) of the same system with $N$ electrons[20]. It was also shown, that when range separation parameters in hybrid functionals are tuned, the curvature is reduced[21] and excited[22] and ground state[23] properties are improved.

In an alternative approach, DE corrections have been developed using an on-site Hubbard U correction in the DFT+U framework, often used in solid-state and transition-metal-containing materials[24, 25] where exact exchange in hybrid functionals incurs higher cost. A correction is carried out within the self-consistent DFT+U approach[26], with a U penalty applied to the localized *d* or *f* subshell. While different methods have been explored to bypass the need for empirical U parameters[27, 28], here we focus on the linear response U, which we refer to as the local curvature. This local curvature is defined as the second derivative of the energy with respect to the removal or addition of charge from a localized subshell $(n_{nl}^I)$ to the rest of the system at constant charge. That is,

$$U_{nl} = \frac{\partial^2 E}{\partial (n_{nl}^I)^2}\bigg|_q \quad (2)$$

where $n_{nl}^I$ is the occupation of the subshell defined by the quantum numbers $n, l$ of atom $I$. The local curvature is calculated with the motivation that its functional form should eliminate the convexity of the energy of an atom as fractional charge is added to or removed from it. The elimination of convexity should subsequently recover piecewise linearity under certain conditions[29, 30].

Although elimination of local curvature and elimination of global curvature have similar aims, the two curvatures represent different properties of the system. In fact, it has been shown that the two curvature values rarely coincide in representative homoleptic transition-metal complexes[31] and instead an inverse relationship has been suggested.[31] We now investigate these two measures of DE in TMCs to determine if similar relationships exist when the ligand field is manipulated by ligand substitution in heteroleptic complexes or by distortion of the metal–ligand bond length in homoleptic complexes. We quantify the extent of ligand additivity[32] in these energetic global and local measures of DE, which has been demonstrated to enable force fields[33] to predict DFT-derived properties. We show that while local and global measures of DE are indeed divergent, concepts of ligand additivity generally hold, suggesting the possibility of quantifying both errors and computing corrections on high-symmetry or small, representative complexes.

We first assembled octahedral complexes where we smoothly vary the ligand field from weak to strong by replacing water or ammonia ligands with carbonyl ligands in heteroleptic complexes. Specifically, we started from a hexa-aqua/hexa-amine complex for our weak-field limit and replaced the ligands one by one with carbonyls until we reached a hexa-carbonyl complex representing the strong-field limit (Figure 1). We then computed properties of heteroleptic complexes between these two limits, with structures obtained from prior work[34]. All stoichiometries were studied except for the *fac* and *mer* complexes with three of each ligand type due to their omission from the complexes studied in prior work[34]. Linear response and curvature calculations were conducted on geometry-optimized structures in both high-spin (HS) and low-spin (LS) states for Mn(II/III) and Fe(II/III). This corresponds to a HS quintet and LS singlet for $d^6$ Fe(II), HS sextet and LS doublet for $d^5$ Mn(II) or Fe(III), and a HS quintet and LS singlet for $d^4$ Mn(III).

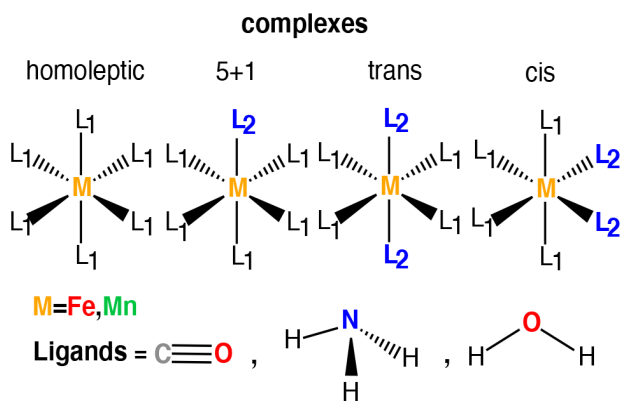

**Figure 1.** Complexes considered in this work. All complexes are octahedral, with metal center M and either one or two types of ligands ($L_1$, $L_2$). For each metal, two ligand combinations were used, one where $L_1$ = CO and $L_2$ = $H_2O$ and one where $L_1$ = CO and $L_2$ = $NH_3$.

To validate our choice of heteroleptic complexes as a way to interpolate between limits of ligand field strength, we analyzed the average metal–ligand bond length as a proxy for the shift in ligand field. As more carbonyl ligands are added to an Fe(II) complex, the LS complex bond lengths decrease, whereas for the HS complex, the average bond length increases (Figure 2). Although the increase in HS bond lengths is surprising, this increase in the HS complex is due to more asymmetry in the bond lengths for this complex in comparison to the LS case (Figure 2 and Supporting Information Figure S1). Importantly, the bond lengths are smoothly varying, verifying the suitability of this model for interpolation between ligand field limits. We next computed linear response U for Fe in a +2.5 oxidation state to balance between the Fe(II) and Fe(III) limits over which global curvature is evaluated. This oxidation state choice was previously shown[31] to provide a roughly average value of those (i.e., Fe(II) and Fe(III)) two limits. We observe a monotonic relationship between nominal ligand field strength and linear response U evaluated for Fe in a +2.5 charge state, with stronger ligand fields leading to higher linear response U values, especially for the LS complex (Figure 2). Throughout the transition from weak to strong ligand field, the HS complex U values are consistently lower than those for the LS state, which is intuitive because hybridization, which is strongly linked to delocalization, is expected to be weaker in the HS state. This effect is significant, as the HS complex U only increases by about 0.3 eV, but the LS complex linear response U increases by nearly 3 eV as the ligand field is varied from weak to strong (Figure 2).

Because the linear response U is an inherently metal-local quantity, it can be expected that changing the metal coordination environment will shift the linear response U value. Indeed, the endpoint values (i.e., of weak field versus strong field) for the Hubbard U match those previously observed[31] in homoleptic complexes, and the heteroleptic complexes studied here provide a smooth interpolation between the endpoint values. The observation that the change in U with increasing number of strong-field ligands is monotonic provides the first evidence that concepts of ligand additivity[32] can be extended to linear response U calculations. Additionally, for some cases we can also evaluate isomer effects. For the case of $Fe(CO)_2(H_2O)_4$, we computed linear response properties for both *cis* and *trans* complexes, and, in agreement with previous work on the sensitivity to fraction of Hartree–Fock (HF) exchange[34], we find that the *trans* $Fe(CO)_2(H_2O)_4$ complexes have linear response U properties that are more similar to the 5+1 complexes than to *cis* 4+2 complexes. Thus, the second *trans* ligand in a *trans* complex has a limited effect on the ligand field strength and thus influences linear response U values in the same fashion as found for HF exchange sensitivity. Although we have focused on linear response U values obtained for $Fe^{2.5+}$, these observations are general to $Fe^{2+}$ and $Fe^{3+}$ complexes (Supporting Information Figure S2).

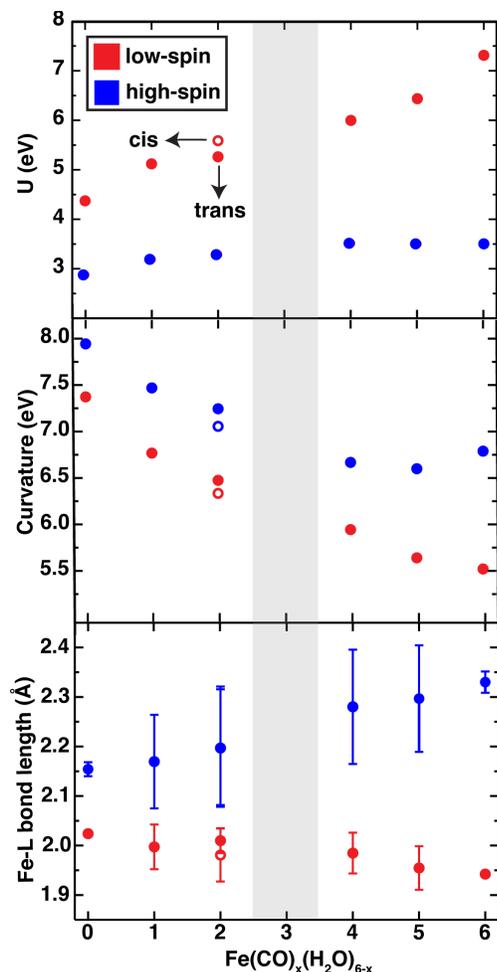

**Figure 2.** Computed properties of Fe(CO)(H$_2$O)$_{6-x}$ complexes: linear response U at oxidation state +2.5 (top panel), global curvature (middle panel) and the average Fe–ligand bond length (lower panel, where the error bars correspond to the standard deviation of the bond lengths), as a function of the number of CO ligands in the transition-metal complex. The *cis* and *trans* configurations of the HS Fe$^{+2.5}$(CO)$_2$(H$_2$O)$_4$ complex in the top panel overlap. The gray shading emphasizes that no isomers were computed at x = 3.

Next, we aimed to determine if the global curvature followed similar trends with ligand field strength. We evaluate the curvature along the fractional charge line between Fe$^{2+}$ and Fe$^{3+}$ as the average curvature between these two states as shown in eq. 1. Consistent with trends observed previously for homoleptic compounds[31], the global curvature decreases with increasing ligand field strength, displaying the opposite trend to the linear response U (Figure 2). For the LS Fe complexes, the curvature decreases by around 2 eV over the range of ligand field strengths accessed by saturating the complex with CO ligands. This effect is likely a result of increased charge delocalization as hybridization increases, i.e., as indicated by shorter average bond lengths in the LS Fe complex with each bonded CO added. Consequently, this leads to a smaller global curvature as bond lengths decrease and ligand field strength increases. Although we observe the same trends on the HS complexes, lower hybridization is present in all complexes, regardless of ligand field, which results in consistently higher curvature values for all high-spin complexes and non-monotonic behavior for addition of more CO ligands from the *trans* $Fe(CO)_4(H_2O)_2$ to the homoleptic hexa-carbonyl complex (Figure 2). The effective hybridization levels off for these complexes, leading to an overall reduction in curvature of only around 1.5 eV for the HS complex over the full range sampled by CO addition (Figure 2).

For the $Fe(CO)_2(H_2O)_4$ complex for which the linear response U had indicated a weaker ligand field for the *trans* complex than the *cis*, we instead observe lower global curvature values consistently for the *cis* case in both LS and HS complexes (Figure 2). In the LS state, the *trans* value appears to align better with the overall linear trend for the complexes containing fewer CO ligands, whereas the *cis* is more consistent with a linear trend for the HS case (Figure 2). In neither case does the *trans* complex global curvature value mimic that of the 5+1 complexes as it did for the linear response U value. Similar results are observed for cases with ammonia ligand substitution and corresponding Mn complexes (Supporting Information Figures S3–S5).

It has been observed that global curvature values and linear response U values are seldom identical in transition-metal complexes[31]. If the two measures were comparable, then application of a single U in DFT+U could simultaneously eliminate both local and global curvature. A secondary concern is that when aiming to compare two different spin states, either local or global measures of curvature might differ for the states being compared, which poses challenges because energies from DFT+U calculations are only comparable when the same value of U is applied to all points. We consider our data set in light of these potential challenges. The linear response U and global curvature are most comparable across both LS and HS states in the strong-field limit, meaning that a single value for the curvature correction may correct both types of delocalization error in a balanced manner for both spin states in this regime.

An inverse relationship between global curvature and the linear response U has previously been shown[31] by comparing homoleptic complexes of strong- and weak-field ligands, i.e, for hexa-carbonyl, hexa-amine and hexa-aqua complexes. As we increase the ligand field strength in this work by adjusting the number of strong-field ligands in a series of heteroleptic complexes, we observe a consistent negative correlation between the two measures of curvature, and that curvature values for intermediate ligand fields ligands reside in between the two homoleptic endpoints. That is, from strong- to weak-field complexes the global curvature decreases while the linear response U increases (Figure 3). Because ammonia and water are comparably weak field ligands, we find that exchanging either ligand with carbonyl has a similar effect on global curvature and linear response U (i.e., local curvature). However, we observe that the relationship between global and local curvature depends strongly on the electron configuration (i.e., spin state) (Figure 3 and Supporting Information Figure S6). Within each spin state (i.e., high- or low-spin), increased ligand field strengths lead to decreased global curvatures and higher linear response U values.

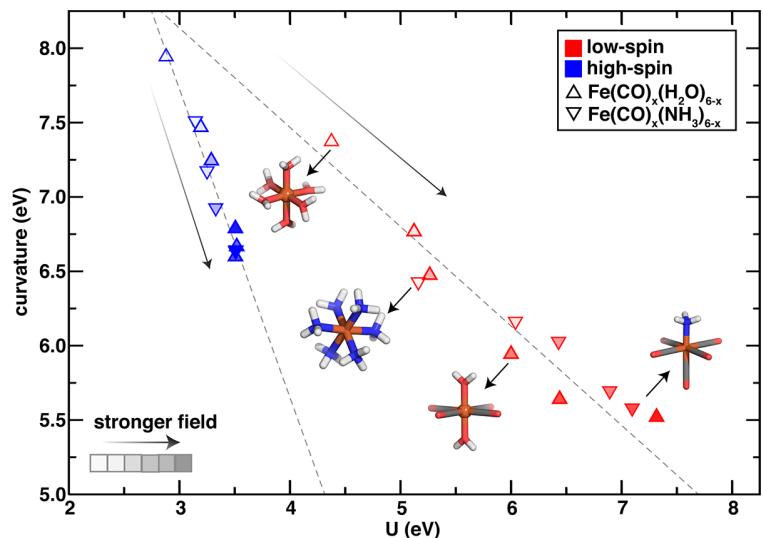

**Figure 3.** The curvature as a function of linear response U values in HS (red) and LS (blue) states for the systems $Fe(CO)_x(H_2O)_{6-x}$ (up triangles) and $Fe(CO)_x(NH_3)_{6-x}$ (down triangles). The global curvature is computed between the +2 and +3 states, whereas the Hubbard U is computed at the +2.5 oxidation state. The transparency of the data points corresponds to the number of strong-field ligands in the complex as indicated in the inset legend in the bottom left corner.

The difference between the spin states is evident from the differences in the slopes of the linear fits for each spin state (Figure 3). In the HS state, the linear response U changes half as fast as the curvature (i.e., slope of -2.07 and $R^2$ of 0.97) as the field strength is tuned. In the LS state, however, the reverse is true, with a change in linear response U that changes twice as much as the curvature (i.e., slope of -0.5 and $R^2$ of 0.77). There is also a greater variation in the data for the HS state. The two linear fits coincide if we extrapolate past the weak-field limits defined by our homoleptic complexes. In both cases, the lines approach a low linear response U (< 3 eV) and a high curvature value (> 8 eV) at the weak-field limit. As ligand field strength increases, the differences in curvature and linear response U values for the different spin states increase (Figure 3). This divergence can be related to differences in the strength of metal–ligand bonding, observed through the bond lengths. The bond lengths for the HS and LS states are closest in value in the case of weak crystal field, and the differences increase with increasing ligand field strength (Figure 2). Thus, the weak-field complexes have more spin-independent bonding covalency in comparison to the stronger-field complexes, as indicated by differences in bond lengths. Our observations do not depend strongly on the metal identity, as we observe similar results for Mn compounds (Supporting Information Figure S5 and Table S1).

We next isolated ligand field strength effects from dependence on the number of atoms or chemical identity of the ligand in evaluating linear response U and the global curvature. To do so, we constructed a homoleptic iron hexa-carbonyl complex, and gradually increased the Fe–C bond length from 2.1 Å to 2.5 Å, both in the HS and LS cases, while holding the C–O bond length fixed to its typical value of 1.12 Å. For the freely optimized homoleptic Fe(III) hexa-carbonyl structures, the Fe–C bond length is 2.3 Å in the HS state and 2.0 Å in the LS state. The Fe–C bonds for the carbonyl ligands in the heteroleptic complexes with weak-field ligands are comparable to their homoleptic counterparts, except for $Fe(CO)(NH_3)_5$ which has a slightly elongated (to ca. 2.5 Å) Fe–C bond (Supporting Information Figure S7). Thus, our sampled distances extend beyond the longest equilibrium bond and are comparable to the shortest strong-field Fe–C bond length. As we decrease the ligand field strength by elongating the Fe–C bond, we also decrease the linear response U for both HS and LS states, consistent with our observations on the heteroleptic compounds (Figure 4). The LS linear response U values are again consistently higher than the HS values (Figure 4). The LS state exhibits a greater variation in Hubbard U with Fe–C distance than the HS state; we observe a 1 eV decrease in U value as the Fe–C bond length increases from 2.2 to 2.5 Å, in comparison to a 0.5 eV increase for the equivalent HS case.

Despite the qualitative consistency between trends observed for heteroleptic complexes and complexes where the Fe–ligand bond length is manipulated, the variation in U with Fe–C bond length variations is smaller than was observed upon substitution of weak-field ligands with carbonyl ligands. Specifically, replacing all water ligands with carbonyls (i.e., x = 0 to 6 in $Fe(CO)_x(H_2O)_{6-x}$) exhibits a change in U that is at least twice as large (Figure 2). The difference is most likely not due to ligand field strength variation. Because these complexes maintain octahedral symmetry, we can use the $d$ orbital splitting as a direct measure of the ligand field strength instead of relying on the proxy of the bond length. We observe this splitting to decrease by ca. 2 eV with Fe–C bond stretching, which is nearly twice that observed with heteroleptic substitutions of carbonyl for weak-field ligands (Supporting Information Figure S8). That is, while changing ligand field through chemical substitution has a significant effect on linear response U, similar changes in ligand field strength by modulating metal–ligand bond has a smaller effect on the linear response U. The linear response U is therefore not only affected by the strength of the ligand field but also by the character of the ligand such as its ability to form $\pi$ bonds or its polarizability.

Variations in global curvature are smaller than variations in the linear response U as Fe–C bonds are stretched (Figure 4). The global curvature in our model system is consistently higher in the HS state than the LS state, in agreement with observations on heteroleptic complexes. Nevertheless, the HS state has an increase in curvature with decreasing ligand field, whereas the LS state curvature is flat or decreases slightly (Figure 4). We observe less than a 0.5 eV increase in curvature for the HS complex with stretched Fe–C bonds compared to a 1.5 eV variation in the set of $Fe(CO)_x(H_2O)_{6-x}$ complexes (Figure 2). This is most likely the result of the frontier orbital energies, which depend on the chemical composition of the entire system, changing less when we only vary bond length and do not vary ligand chemistry. In the LS case in particular, decreasing the ligand field by stretching the Fe–C bond in our homoleptic model complex does not result in the same increase in the global curvature that we observed when we decreased the ligand field by replacing strong-field ligands with weak-field ligands. These differences can be rationalized by noting that global curvature is more sensitive to the nature of the ligands in the complex than to the strength of the metal–ligand bond.

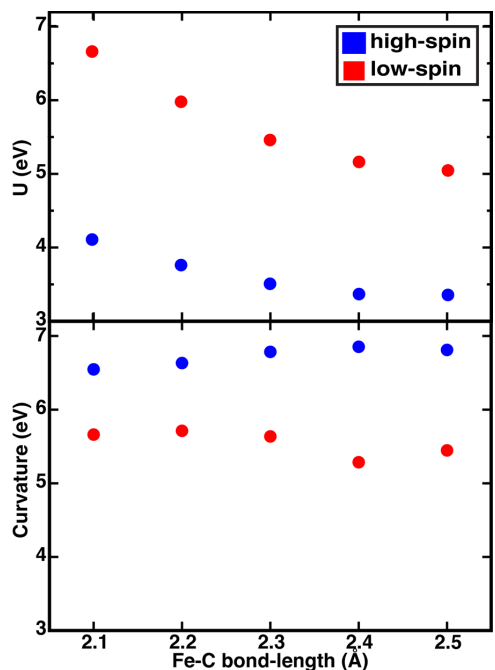

**Figure 4.** Linear response U for Fe in the +2.5 oxidation state (upper panel) and the global curvature (lower panel) as a function of Fe–C bond length in the LS state (red data points) and the HS state (blue data points).

In summary, we investigated the effect of the ligand field on measures of delocalization error (i.e., local and global curvature) in model systems. We both systematically substituted weak-field ligands with strong-field ligands and altered the Fe–ligand bond in a homoleptic hexa-carbonyl complex. We confirmed that ligand field has a strong effect on the two measures. An increase in ligand field strength results in an increase in linear response U and a decrease in the global curvature, and this relationship is dependent on spin state. The effect is more pronounced in the LS cases, likely due to shorter metal–ligand bonds in the LS transition-metal complex, which corresponds to more hybridization of the frontier orbitals and therefore a larger effect on the charge distribution in the systems. Importantly, global and local curvatures are seldom the same in any homoleptic or heteroleptic complex due to their inverse behavior, but they are most similar for intermediate field strength LS complexes. We confirmed that ligand additivity holds in measures of delocalization error. Because we have shown that ligand additivity relationships are applicable to delocalization error, this principle can be useful for predicting and correcting errors in heteroleptic complexes from homoleptic endpoints. Such a strategy could be used for eliminating delocalization errors in DFT based on benchmarks from delocalization-error-free correlated wavefunction theory or corrections (i.e., curvature values) from homoleptic complexes.

*Computational Details.* We used optimized geometries from prior studies[34] for all Fe complexes and for the low-spin Mn complexes discussed in the work. The set contains homoleptic complexes (i.e., hexa-aqua, hexa-ammine, or hexa-carbonyl) in addition to heteroleptic mononuclear octahedral complexes made from combinations of strong-field CO and either $H_2O$ or $NH_3$ weak-field ligands (Figure 1). The complexes have the formula $M(CO)_x(L)_{6-x}$ where L = $H_2O$ or $NH_3$ and M = Mn or Fe. The x = 2 cases can be arranged in *cis* or *trans* configurations. Only the x = 4 *trans* configuration was generated in previous work and is thus the only isomer used here. Similarly, we do not include x = 3 cases (i.e., in either *facial* or *meridional* configurations) which were not computed in prior studies[34]. Optimizations in the prior work were obtained using a development version of TeraChem,[35, 36] with the B3LYP[37] GGA hybrid functional and the LACVP* composite atom-centered basis set[34]. The LACVP* basis set consists of the LANL2DZ effective core potential on the metal and 6-31G* on the remaining atoms. Due to challenges converging PBE[38] semi-local GGA exchange-correlation (xc) functional single-point energy calculations on these B3LYP structures for high-spin Mn complexes for the present work, we re-optimized these complexes using the plane-wave periodic boundary condition code Quantum-ESPRESSO v6.7[39] with the PBE xc functional starting from B3LYP/LACVP*-optimized LS Mn(II) compounds.

We used molSimplify[40, 41] to construct octahedral structures with uniform Fe–C bond length that varied from 2.1 Å to 2.5 Å in the HS and LS cases in increments of 0.1 Å. All structures, including both from prior work and the reoptimized structures, are provided in the Supporting Information .zip file.

All single-point density functional theory (DFT) calculations were carried out with Quantum-ESPRESSO v6.7 using the PBE GGA xc functional. Ultrasoft pseudopotentials[42] (USPPs) obtained from the Quantum-ESPRESSO website[43] were employed, which have semi-core $3s$ and $3p$ states included in the valence for both Fe and Mn. We used a plane-wave energy cutoff of 30 Ry for the wavefunction and 300 Ry for the charge density. With the exception of low-spin (LS) $Fe(CO)_6$, all transition-metal complexes were placed in a cubic box of 18.5 Å to ensure sufficient vacuum. For LS $Fe(CO)_6$, we increased the box size to 26.5 Å to test its effect on the results. The Martyna–Tuckerman scheme[44] was employed to eliminate periodic image effects for all calculations. Consistent with prior work, at least 15 unoccupied bands were included in all calculations.

To study delocalization error, we evaluated or approximated (i.e., from eigenvalues) fractional-charge properties between the +2 and +3 metal oxidation states in this work for global curvature and linear response U. We chose high-spin and low-spin conventions in accordance with prior work as follows. The Mn(II)/Mn(III) spin multiplicity decreases from sextet to quintet in the high-spin configuration and from doublet to singlet in the low-spin case, as a majority-spin electron is removed (Supporting Information Table S2). The Fe(II)/Fe(III) spin multiplicity increases from quintet to sextet and from singlet to doublet in high- and low-spin configurations as a minority-spin electron is removed from the system (Supporting Information Table S2). The HOMO and LUMO, which are used in the average curvature expression, were chosen as the levels from which an electron is removed or to which an electron is added according to these conventions, respectively.

To compute global curvature, we used the average curvature expression as in eq. 1 for the global curvature[20] because calculations based on fractional charge differ only by a modest rigid shift of around 0.5 eV (Supporting Information Figure S9). To compute local curvature, linear response calculations were carried out within Quantum-ESPRESSO v6.7 using a rigid potential shift ranging from -0.08 eV to 0.08 eV in increments of 0.02 eV. Fractional charge linear response U calculations were carried out in the same manner, where the minority-spin electron in Fe and majority-spin electron in Mn complexes were manually set to 0.5 using the "from input" command when defining occupations in Quantum-ESPRESSO. The linear response U and global curvature calculations were carried out in most cases using the complex structure for the M(III) oxidation state and the corresponding HS or LS state. The sole exception to this was LS Mn for which Mn(III) structures could not be converged. In that case, they were instead carried out using the complex structure for the Mn(II) oxidation state.

ASSOCIATED CONTENT

**Supporting Information.** The following files are available free of charge.

Linear response U at different oxidation states, global and local curvatures of Fe complexes with ammonia and carbonyl ligands, global and local curvatures of Mn complexes, global curvature as function of local curvature for Mn complexes, linear fits to global vs local curvature for the Mn case, Mean Fe–C bond length for Fe complexes with ammonia ligands combined with carbonyl and for complexes with water ligands combined with carbonyl, d-splitting energy in Fe hexa-carbonyl complexes with varying bond lengths and in the freely optimized Fe complexes with ammonia and carbonyl ligand combinations, Formal oxidation states and corresponding spin states discussed in this work, comparison of global curvature calculated in different methods. (PDF)

Optimized structures of Fe(II), Fe(III) and Mn(II), Mn(III) structures with ammonia, carbonyl and water carbonyl ligand combinations, Fe hexa-carbonyl structures with different Fe–C bond lengths (ZIP)

AUTHOR INFORMATION

**Notes**

The authors declare no competing financial interests.

ACKNOWLEDGMENT

The authors acknowledge primary support by the Department of Energy under grant number DE-SC0018096 (for Y.C. and A.B.). This material was also based upon work supported by the Department of Energy, National Nuclear Security Administration under Award Number DE-NA0003965 (for H.J.K.). Generation of the datasets in this work were supported by the Office of Naval Research under grant numbers N00014-18-1-2434 and N00014-20-1-2150 (to A.N. and A.B.). A.N. was partially supported by a National Science Foundation Graduate Research Fellowship under Grant #1122374. This work was also carried out in part using computational resources from the Extreme Science and Engineering Discovery Environment (XSEDE), which is supported by National Science Foundation grant number ACI-1548562. H.J.K. holds a Career Award at the Scientific Interface from the Burroughs Wellcome Fund, an AAAS Marion Milligan Mason Award, and an Alfred P. Sloan award in Chemistry, which supported this work. The authors thank Isuru Ariyarathna, Yeongsu Cho, Chenru Duan and Adam H. Steeves for providing a critical reading of the manuscript.

REFERENCES

1.	Niu, S.; Hall, M. B., Theoretical Studies on Reactions of Transition-Metal Complexes. *Chemical Reviews* **2000,** *100* (2), 353-406.
2.	Haunschild, R.;  Henderson, T. M.;  Jiménez-Hoyos, C. A.; Scuseria, G. E., Many-electron self-interaction and spin polarization errors in local hybrid density functionals. *The Journal of Chemical Physics* **2010,** *133* (13), 134116.
3.	Mori-Sánchez, P.;  Cohen, A. J.; Yang, W., Many-electron self-interaction error in approximate density functionals. *The Journal of Chemical Physics* **2006,** *125* (20), 201102.
4.	Ruzsinszky, A.;  Perdew, J. P.;  Csonka, G. I.;  Vydrov, O. A.; Scuseria, G. E., Density functionals that are one- and two- are not always many-electron self-interaction-free, as shown for H2+, He2+, LiH+, and Ne2+. *The Journal of Chemical Physics* **2007,** *126* (10), 104102.
5.	Cohen, A. J.;  Mori-Sánchez, P.; Yang, W., Insights into Current Limitations of Density Functional Theory. *Science* **2008,** *321* (5890), 792-794.
6.	Schmidt, T.; Kümmel, S., One- and many-electron self-interaction error in local and global hybrid functionals. *Physical Review B* **2016,** *93* (16), 165120.
7.	Ruzsinszky, A.;  Perdew, J. P.;  Csonka, G. I.;  Vydrov, O. A.; Scuseria, G. E., Spurious fractional charge on dissociated atoms: Pervasive and resilient self-interaction error of common density functionals. *The Journal of Chemical Physics* **2006,** *125* (19), 194112.
8.	Dutoi, A. D.; Head-Gordon, M., Self-interaction error of local density functionals for alkali–halide dissociation. *Chemical Physics Letters* **2006,** *422* (1), 230-233.
9.	Bally, T.; Sastry, G. N., Incorrect Dissociation Behavior of Radical Ions in Density Functional Calculations. *The Journal of Physical Chemistry A* **1997,** *101* (43), 7923-7925.
10.	Zhang, Y.; Yang, W., A challenge for density functionals: Self-interaction error increases for systems with a noninteger number of electrons. *The Journal of Chemical Physics* **1998,** *109* (7), 2604-2608.
11.	Johnson, B. G.;  Gonzales, C. A.;  Gill, P. M. W.; Pople, J. A., A density functional study of the simplest hydrogen abstraction reaction. Effect of self-interaction correction. *Chemical Physics Letters* **1994,** *221* (1), 100-108.
12.	Mori-Sánchez, P.;  Cohen, A. J.; Yang, W., Localization and Delocalization Errors in Density Functional Theory and Implications for Band-Gap Prediction. *Physical Review Letters* **2008,** *100* (14), 146401.
13.	Cohen, A. J.;  Mori-Sánchez, P.; Yang, W., Fractional charge perspective on the band gap in density-functional theory. *Physical Review B* **2008,** *77* (11), 115123.
14.	Tozer, D. J.; De Proft, F., Computation of the Hardness and the Problem of Negative Electron Affinities in Density Functional Theory. *The Journal of Physical Chemistry A* **2005,** *109* (39), 8923-8929.
15.	Teale, A. M.;  De Proft, F.; Tozer, D. J., Orbital energies and negative electron affinities from density functional theory: Insight from the integer discontinuity. *The Journal of Chemical Physics* **2008,** *129* (4), 044110.
16.	Peach, M. J. G.;  Teale, A. M.;  Helgaker, T.; Tozer, D. J., Fractional Electron Loss in Approximate DFT and Hartree–Fock Theory. *Journal of Chemical Theory and Computation* **2015,** *11* (11), 5262-5268.
17.	Zheng, X.;  Liu, M.;  Johnson, E. R.;  Contreras-García, J.; Yang, W., Delocalization error of density-functional approximations: A distinct manifestation in hydrogen molecular chains. *The Journal of Chemical Physics* **2012,** *137* (21), 214106.

18.    Perdew, J. P.; Parr, R. G.; Levy, M.; Balduz, J. L., Density-Functional Theory for Fractional Particle Number: Derivative Discontinuities of the Energy. *Physical Review Letters* **1982,** *49* (23), 1691-1694.
19.    Koopmans, T., Über die Zuordnung von Wellenfunktionen und Eigenwerten zu den einzelnen Elektronen eines Atoms. *Physica* **1934,** *1* (1), 104-113.
20.    Stein, T.; Autschbach, J.; Govind, N.; Kronik, L.; Baer, R., Curvature and Frontier Orbital Energies in Density Functional Theory. *J Phys Chem Lett* **2012,** *3* (24), 3740-4.
21.    Baer, R.; Livshits, E.; Salzner, U., Tuned Range-Separated Hybrids in Density Functional Theory. *Annual Review of Physical Chemistry* **2010,** *61* (1), 85-109.
22.    Refaely-Abramson, S.; Baer, R.; Kronik, L., Fundamental and excitation gaps in molecules of relevance for organic photovoltaics from an optimally tuned range-separated hybrid functional. *Physical Review B* **2011,** *84* (7), 075144.
23.    Srebro, M.; Autschbach, J., Does a Molecule-Specific Density Functional Give an Accurate Electron Density? The Challenging Case of the CuCl Electric Field Gradient. *The Journal of Physical Chemistry Letters* **2012,** *3* (5), 576-581.
24.    Xu, Z.; Rossmeisl, J.; Kitchin, J. R., A Linear Response DFT+U Study of Trends in the Oxygen Evolution Activity of Transition Metal Rutile Dioxides. *The Journal of Physical Chemistry C* **2015,** *119* (9), 4827-4833.
25.    Xu, Z.; Joshi, Y. V.; Raman, S.; Kitchin, J. R., Accurate electronic and chemical properties of 3d transition metal oxides using a calculated linear response U and a DFT + U(V) method. *The Journal of Chemical Physics* **2015,** *142* (14), 144701.
26.    Cococcioni, M.; de Gironcoli, S., Linear response approach to the calculation of the effective interaction parameters in the LDA+U method. *Physical Review B* **2005,** *71* (3).
27.    Mosey, N. J.; Carter, E. A., Ab initio evaluation of Coulomb and exchange parameters for $\mathrm{DFT}+\mathrm{U}$ calculations. *Physical Review B* **2007,** *76* (15), 155123.
28.    Mosey, N. J.; Liao, P.; Carter, E. A., Rotationally invariant ab initio evaluation of Coulomb and exchange parameters for DFT+U calculations. *The Journal of Chemical Physics* **2008,** *129* (1), 014103.
29.    Himmetoglu, B.; Floris, A.; de Gironcoli, S.; Cococcioni, M., Hubbard-corrected DFT energy functionals: The LDA+U description of correlated systems. *International Journal of Quantum Chemistry* **2014,** *114* (1), 14-49.
30.    Anisimov, V. I.; Solovyev, I. V.; Korotin, M. A.; Czyżyk, M. T.; Sawatzky, G. A., Density-functional theory and NiO photoemission spectra. *Physical Review B* **1993,** *48* (23), 16929-16934.
31.    Zhao, Q.; Ioannidis, E. I.; Kulik, H. J., Global and local curvature in density functional theory. *Journal of Chemical Physics* **2016,** *145*, 054109.
32.    Glerup, J.; Moensted, O.; Schaeffer, C. E., Nonadditive and additive ligand fields and spectrochemical series arising from ligand field parameterization schemes. Pyridine as a nonlinearly ligating .pi.-back-bonding ligand toward chromium(III). *Inorganic Chemistry* **1976,** *15* (6), 1399-1407.
33.    Deeth, R. J.; Foulis, D. L.; Williams-Hubbard, B. J., Molecular mechanics for multiple spin states of transition metal complexes. *Dalton Transactions* **2003,** (20), 3949-3955.
34.    Nandy, A.; Chu, D. B. K.; Harper, D. R.; Duan, C.; Arunachalam, N.; Cytter, Y.; Kulik, H. J., Large-scale comparison of 3d and 4d transition metal complexes illuminates the reduced effect of exchange on second-row spin-state energetics. *Physical Chemistry Chemical Physics* **2020,** *22* (34), 19326-19341.

35. Ufimtsev, I. S.; Martinez, T. J., Quantum chemistry on graphical processing units. 3. Analytical energy gradients, geometry optimization, and first principles molecular dynamics. *Journal of Chemical Theory and Computation* **2009,** *5* (10), 2619-2628.
36. Petachem. http://www.petachem.com. (accessed April 1, 2020).
37. Becke, A. D., Density-functional thermochemistry. III. The role of exact exchange. *Journal of Chemical Physics* **1993,** *98* (7), 5648-5652.
38. Perdew, J. P.; Burke, K.; Ernzerhof, M., Generalized gradient approximation made simple. *Physical review letters* **1996,** *77* (18), 3865.
39. Giannozzi, P.; Andreussi, O.; Brumme, T.; Bunau, O.; Buongiorno Nardelli, M.; Calandra, M.; Car, R.; Cavazzoni, C.; Ceresoli, D.; Cococcioni, M.; Colonna, N.; Carnimeo, I.; Dal Corso, A.; de Gironcoli, S.; Delugas, P.; DiStasio, R. A.; Ferretti, A.; Floris, A.; Fratesi, G.; Fugallo, G.; Gebauer, R.; Gerstmann, U.; Giustino, F.; Gorni, T.; Jia, J.; Kawamura, M.; Ko, H. Y.; Kokalj, A.; Küçükbenli, E.; Lazzeri, M.; Marsili, M.; Marzari, N.; Mauri, F.; Nguyen, N. L.; Nguyen, H. V.; Otero-de-la-Roza, A.; Paulatto, L.; Poncé, S.; Rocca, D.; Sabatini, R.; Santra, B.; Schlipf, M.; Seitsonen, A. P.; Smogunov, A.; Timrov, I.; Thonhauser, T.; Umari, P.; Vast, N.; Wu, X.; Baroni, S., Advanced capabilities for materials modelling with Quantum ESPRESSO. *Journal of Physics: Condensed Matter* **2017,** *29* (46), 465901.
40. Ioannidis, E. I.; Kulik, H. J., Towards quantifying the role of exact exchange in predictions of transition metal complex properties. *The Journal of Chemical Physics* **2015,** *143* (3), 034104.
41. Nandy, A.; Duan, C.; Janet, J. P.; Gugler, S.; Kulik, H. J., Strategies and Software for Machine Learning Accelerated Discovery in Transition Metal Chemistry. *Industrial & Engineering Chemistry Research* **2018,** *57* (42), 13973-13986.
42. Vanderbilt, D., Soft self-consistent pseudopotentials in a generalized eigenvalue formalism. *Physical Review B* **1990,** *41* (11), 7892-7895.
43. https://www.quantum-espresso.org/pseudopotentials.
44. Martyna, G. J.; Tuckerman, M. E., A reciprocal space based method for treating long range interactions in ab initio and force-field-based calculations in clusters. *The Journal of Chemical Physics* **1999,** *110* (6), 2810-2821.

# Supporting Information for

*Ligand Additivity and Divergent Trends in Two Types of Delocalization Errors from Approximate Density Functional Theory*

Yael Cytter [1], Aditya Nandy[1,2], Akash Bajaj[1,3], and Heather J. Kulik[1,*]

[1]Department of Chemical Engineering, Massachusetts Institute of Technology, Cambridge, MA 02139, USA

[2]Department of Chemistry, Massachusetts Institute of Technology, Cambridge, MA 02139, USA

[3]Department of Materials Science and Engineering, Massachusetts Institute of Technology, Cambridge, MA 02139, USA

*email: hjkulik@mit.edu, phone: 617-253-4584

**Contents**

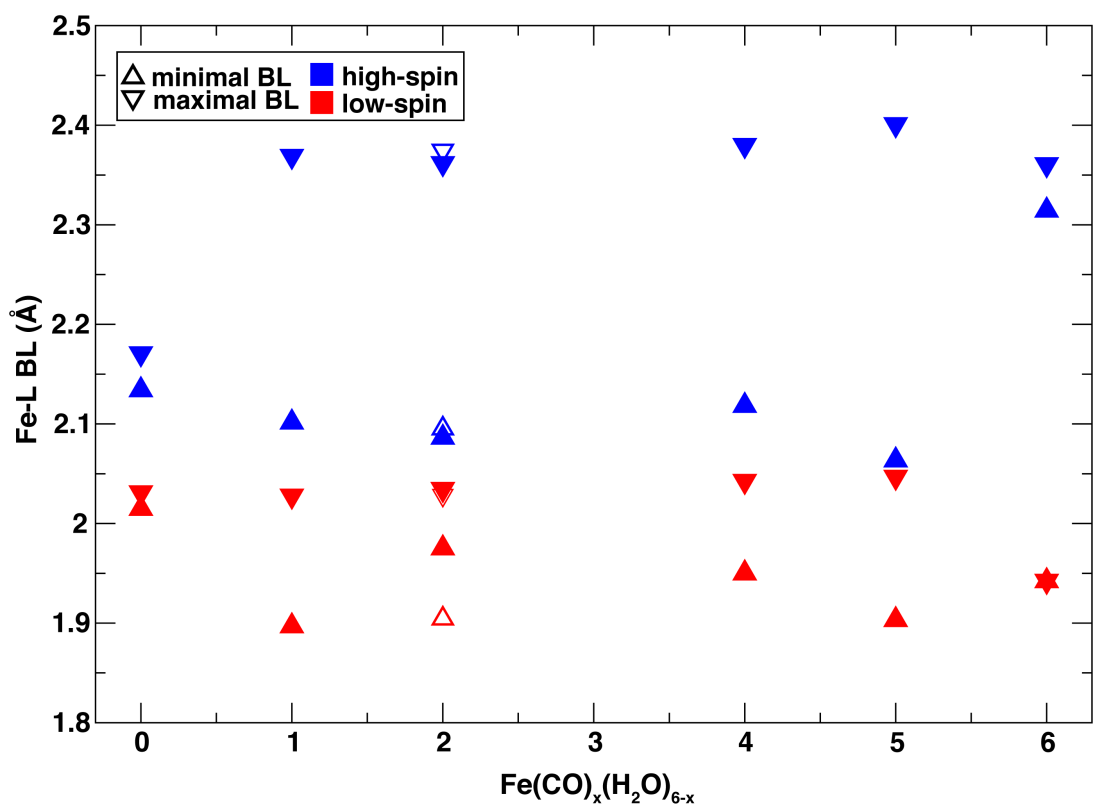

**Figure S1.** The maximal (triangle down) and minimal (triangle up) Fe–ligand bond lengths for the Fe(CO)$_x$(H$_2$O)$_{6-x}$ complexes, in the HS (blue) and LS (red) states. The empty markers correspond the *cis* configuration of the Fe(CO)$_x$(H$_2$O)$_{6-x}$ complex and the full ones correspond to the *trans* configuration.

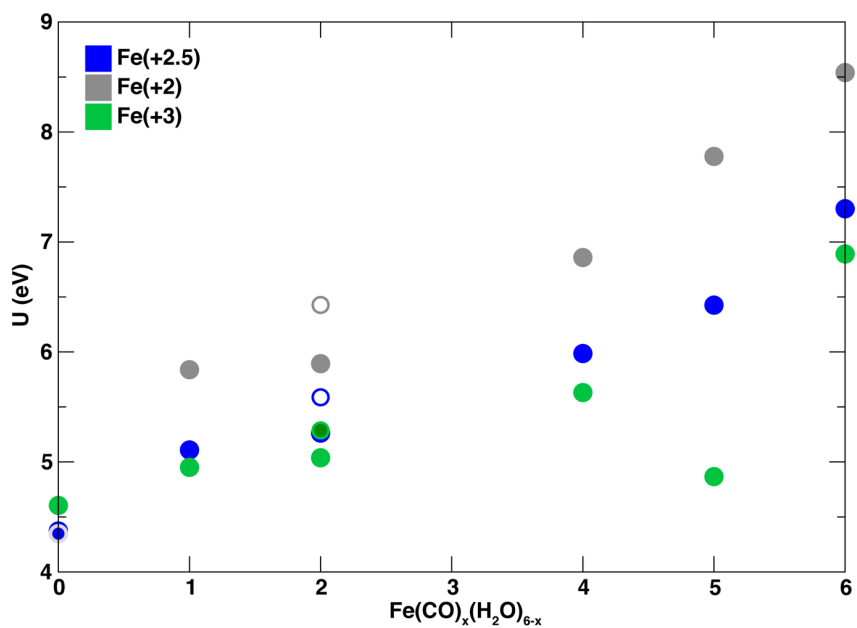

**Figure S2**. Linear response U for $Fe(CO)_x(H_2O)_{6-x}$ complexes in LS states, for oxidation state +2 (gray), +2.5 (blue) and +3 (green). Empty markers correspond to the *cis* configuration of $Fe(CO)_2(H_2O)_4$. Linear response U in the $Fe^{+3}(CO)_5(H_2O)$ complex is an outlier to the overall trend observed possibly due to differences in the converged electronic state.

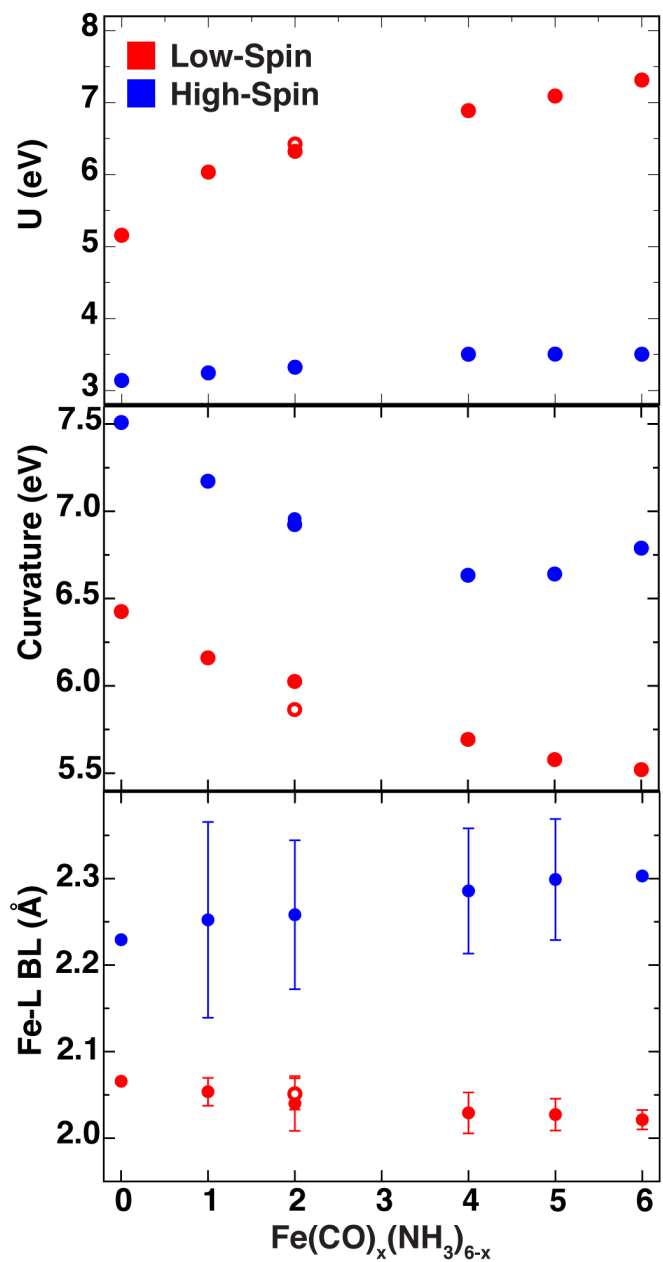

**Figure S3.** (top) Linear response U for oxidation state +2.5 complexes, (middle) global curvature (bottom) and average $Fe^{2+}$–ligand bond length in Å for variation from hexa-ammine to hexa-carbonyl iron complexes. Error bars correspond to the standard deviation of the bond length. All data are shown as a function of the number of carbonyl ligands in an octahedral Fe complex with carbonyl and ammonia ligands, in the high-spin (blue) and low-spin (red) states. The empty circles correspond to the *cis* configuration of $Fe(CO)_2(NH_3)_4$. High-spin cis and trans data points overlap in all panels.

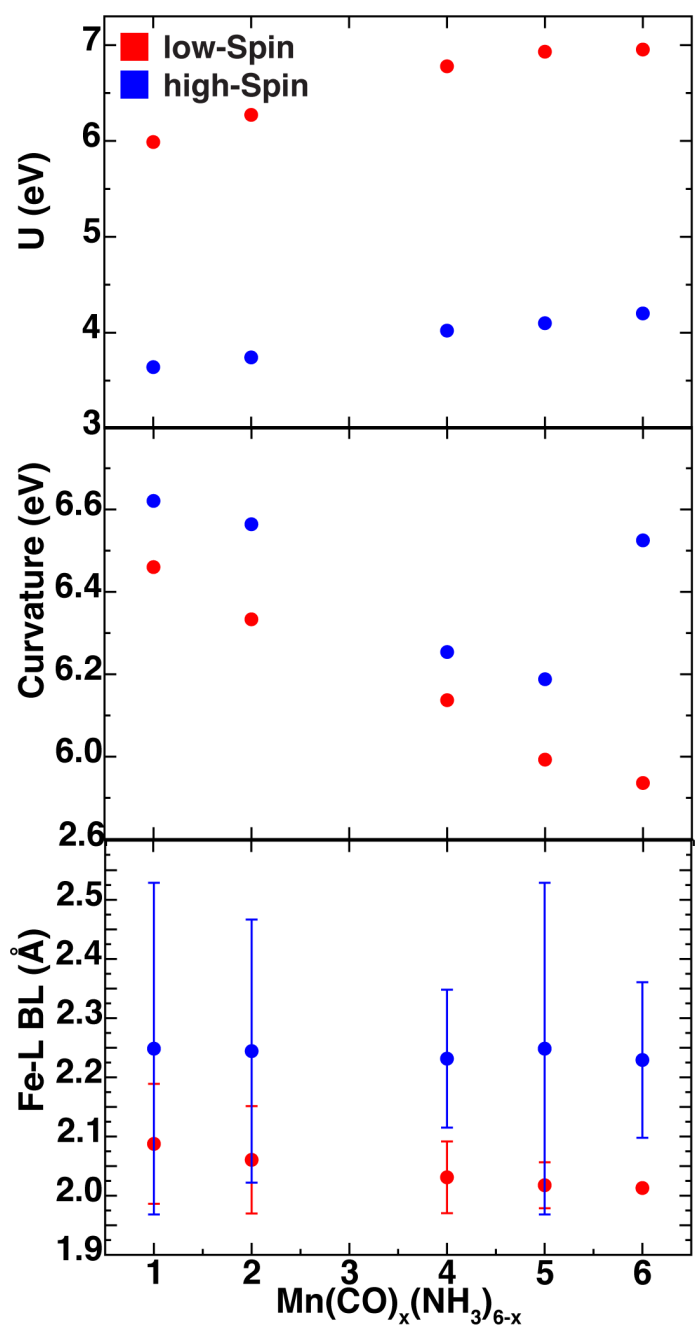

**Figure S4.** (top) Linear response U for oxidation state +2.5 complexes, (middle) global curvature (bottom) and average Mn–ligand bond length in Å for variation from hexa-ammine to hexa-carbonyl manganese complexes. Error bars correspond to the standard deviation of the bond length. All data are shown as a function of the number of carbonyl ligands in an octahedral Mn complex with carbonyl and ammonia ligands, in the high-spin (blue) and low-spin (red) states. Variation in the curvature is in this complex is small compared to other systems: 0.5 eV here vs. 1 eV in the corresponding complex with Fe for example in Figure S2. The single-point calculation for the $Mn^{+3}(NH_3)_6$ endpoint did not converge. As a result, the global curvature could not be calculated for this point. Therefore, this complex was omitted from the comparison. In addition, the *cis* Mn(CO)2(NH3)4 complex is not displayed because geometry optimizations did not succeed.

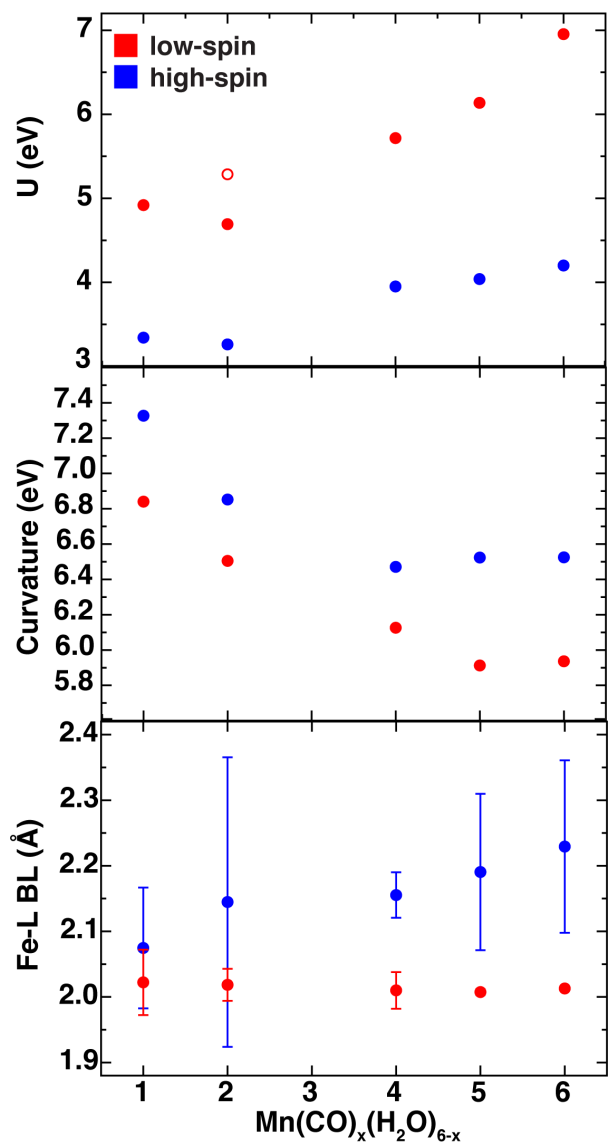

**Figure S5.** (top) Linear response U for oxidation state +2.5 complexes, (middle) global curvature (bottom) and average Mn–ligand bond length in Å for variation from hexa-aqua to hexa-carbonyl manganese complexes. Error bars correspond to the standard deviation of the bond length. All data are shown as a function of the number of carbonyl ligands in an octahedral Mn complex with carbonyl and water ligands, in the high-spin (blue) and low-spin (red) states. The empty circles correspond to the *cis* configuration of Mn(CO)2(H2O)4. The geometry optimization for Mn hexa-aqua complex and the trans configuration of Mn(CO)2(H2O)4 did not converge and are therefore not displayed in the plot. In addition, the single-point calculation for *trans* configuration of the Mn$^{+3}$(CO)$_2$(H$_2$O)$_4$ endpoint did not converge and therefore the curvature for this structure is not displayed.

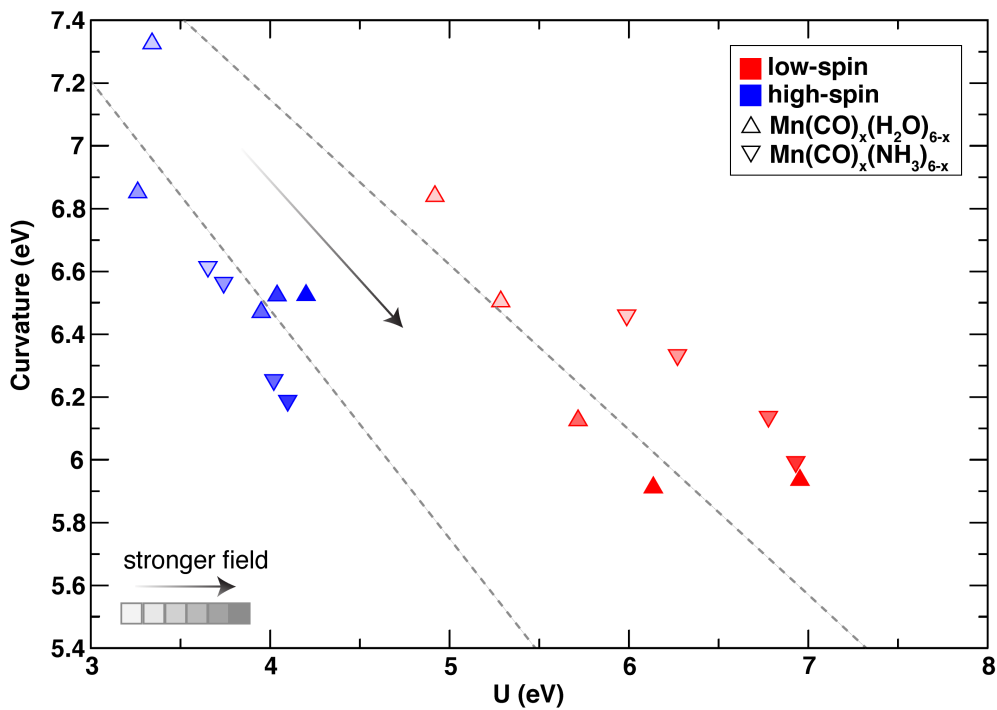

**Figure S6**. Linear response U values vs global curvature values (both in eV) for the HS (blue) and LS (red) states for the complexes Mn(CO)$_x$(H$_2$O)$_{6-x}$ (up triangles) and Mn(CO)$_x$(NH$_3$)$_{6-x}$ (down triangles). The transparency of the data points corresponds to the ligand field where more opaque symbol filling refers to stronger ligand field i.e, more carbonyl ligands and fewer weak-field ammonia or water ligands. The dotted lines are linear fits, performed separately for each spin state (properties shown in Table S2). For simplicity, only the trans structure of the Fe(CO)$_2$(H$_2$O)$_4$ and Fe(CO)$_2$(NH$_3$)$_4$ complexes are shown in the figure. Mn hexa-aqua is not displayed in this figure due to convergence issues in the geometry optimization of the complex. Mn hexa-ammine is not shown either because the single-point calculation of the +3-oxidation state did not succeed.

**Table S1.** Slope and goodness of fit for linear fits to curvature as a function of linear response U in Mn complexes, in HS and LS states.

|  | HS | LS |
| --- | --- | --- |
| Slope | -0.729 | -0.351 |
| $R^2$ | 0.61 | 0.68 |

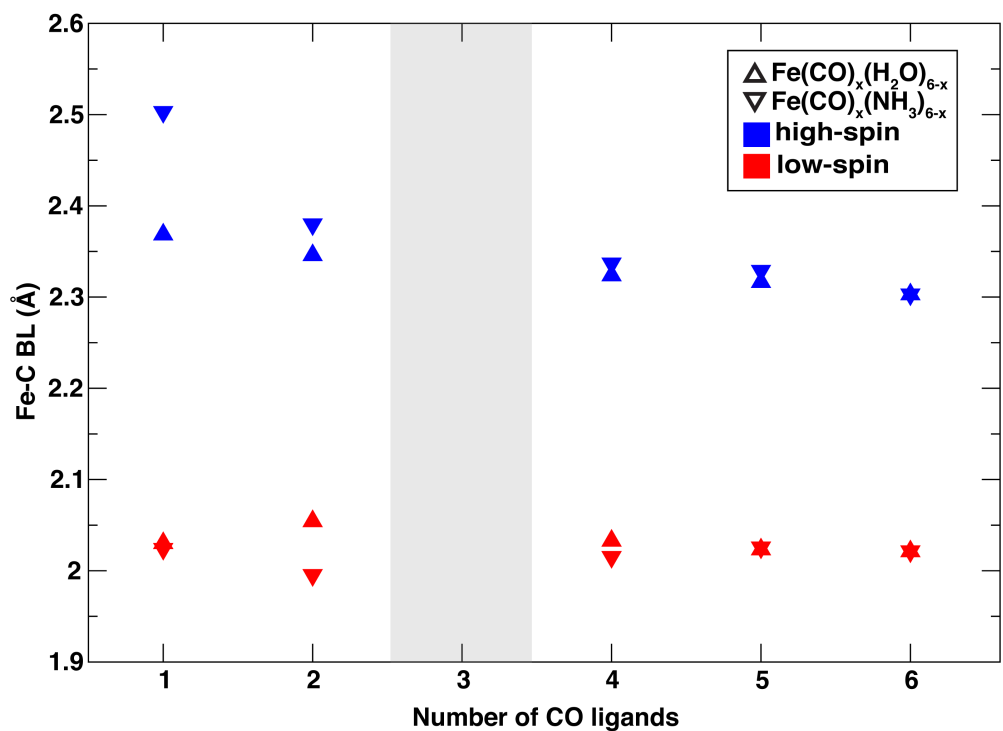

**Figure S7.** Fe–C mean bond length as a function of the number of CO ligands in the complex for $Fe(CO)_x(H_2O)_{(6-x)}$ complexes (up triangles) and $Fe(CO)_x(NH_3)_{(6-x)}$ (down triangles) in HS (blue) and LS (red) states. Error bars are not included in the figure because in the majority of cases they proved to be smaller than the marker size. The highest observed standard deviation is 0.02 Å.

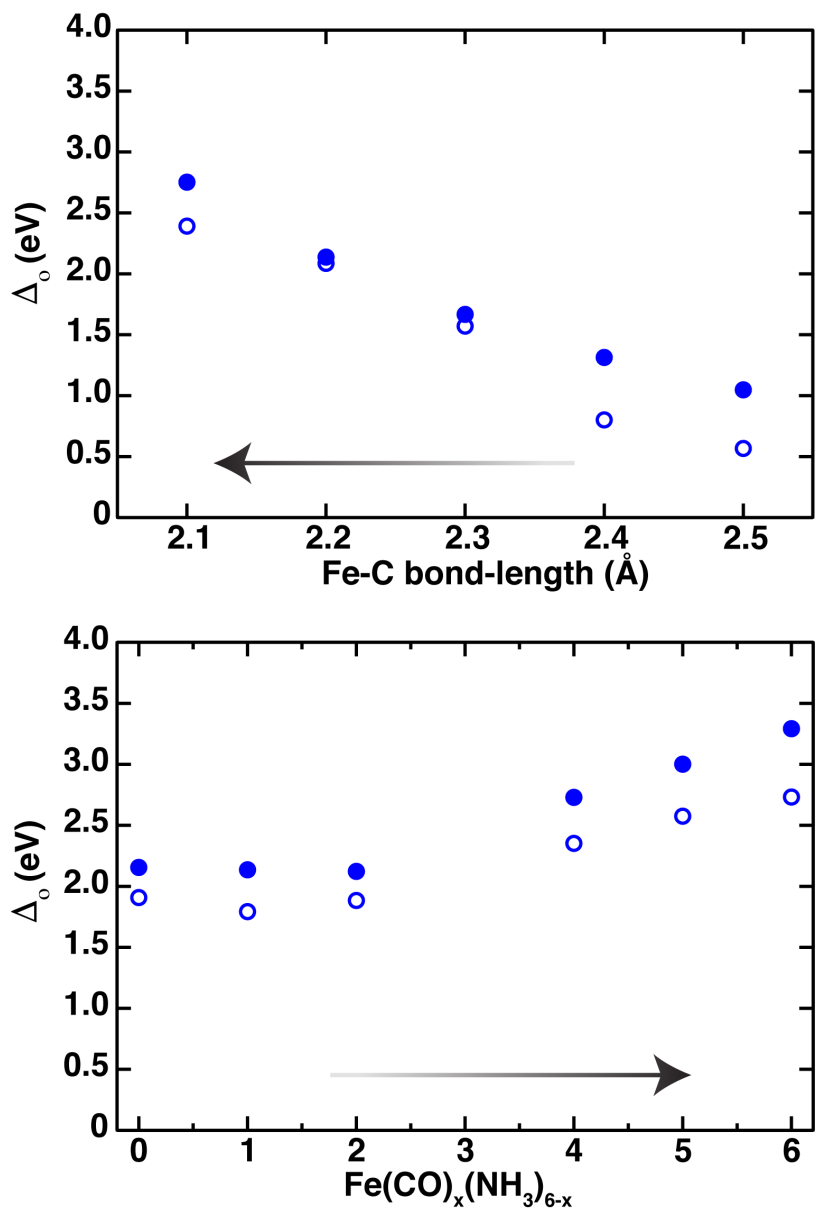

**Figure S8.** The d-splitting energy ($\Delta_O$) as a function of the Fe–C bond length in the LS hexa-carbonyl Fe(II) and Fe(III) complex (top figure) and the d-splitting energy as a function the number of $NH_3$ ligands in $Fe(CO)_x(NH_3)_{6-x}$ in the LS complex (bottom figure). The d-splitting energy was calculated from the eigenvalues in the LS state, i.e., the energy of the LUMO minus the energy of the HOMO. The filled circles correspond to Fe in a +2 oxidation state, and the empty circles correspond to Fe in a +3 oxidation state, the two endpoints by which the average curvature is calculated. The arrows in the figures represents the direction of increasing field strength.

**Table S2.** Formal oxidation and spin multiplicities (2*S*+1) considered in this work. The first number is the multiplicity of the low-spin state and the second number is the multiplicity of the high-spin state. For each metal, global curvatures were obtained using the oxidation states as the energy deviation end points, both in high-spin and low-spin states. Linear response calculations were performed on each oxidation state in both spin states.

| metal | oxidation state | spin multiplicities |
|---|---|---|
| Fe | +2 | 1, 5 |
| Fe | +3 | 2, 6 |
| Mn | +2 | 2, 6 |
| Mn | +3 | 1, 5 |

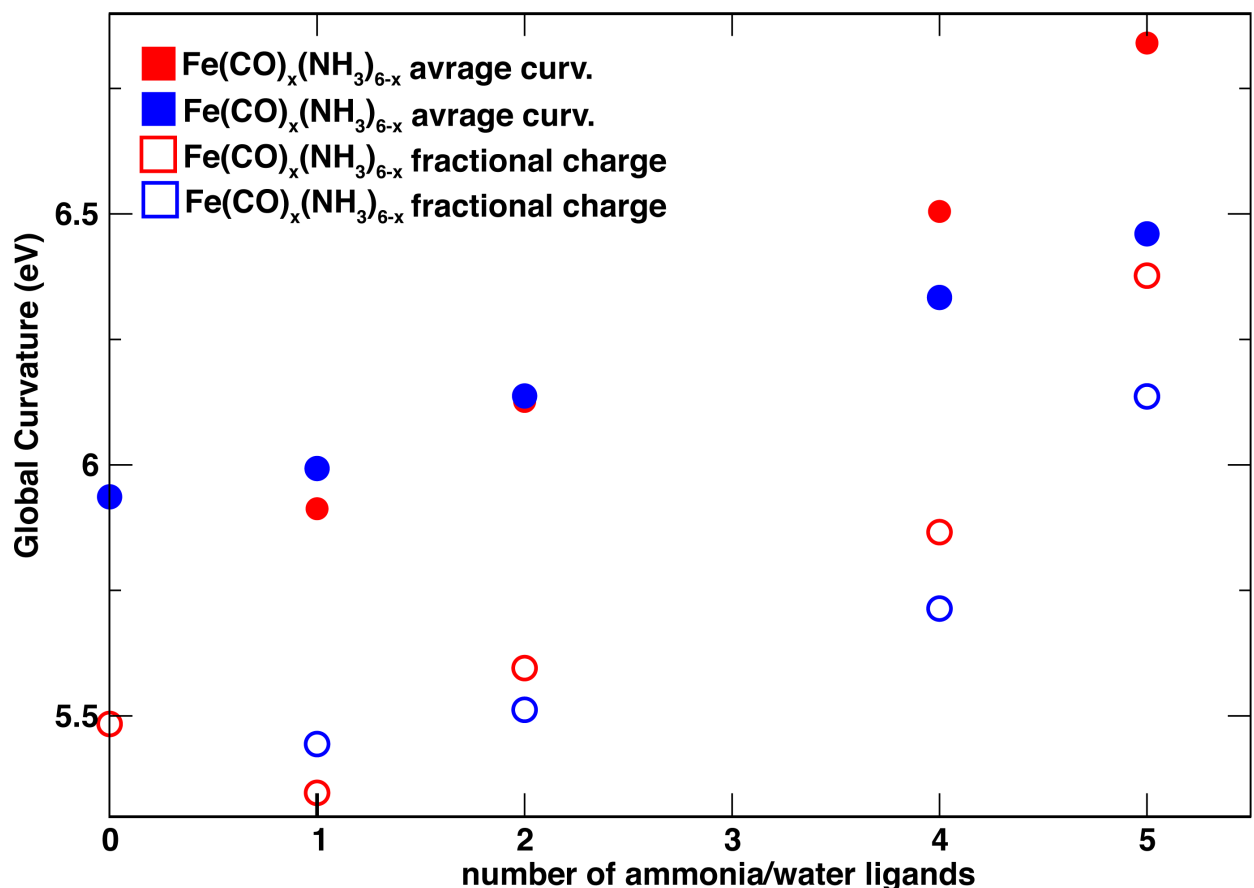

**Figure S9.** Global curvature $\frac{\partial^2 E}{\partial q^2}$ as a function of the number of ammonia (blue) or water (red) ligands in a $Fe(CO)_{6-x}(H_2O)_x$ or $Fe(CO)_{6-x}(NH_3)_x$ complex. Full markers correspond to the average curvatures, $\epsilon_N^{HOMO} - \epsilon_{N-1}^{LUMO}$, and empty markers correspond to curvatures obtained from the second derivative of a second order polynomial fit to fractional charge.